\documentclass[aps,prl,reprint]{revtex4-1}
\usepackage[dvips]{graphicx}
\usepackage{amsmath}

\begin{document}

\title{Quantum efficiency of III-Nitride emitters: evidence for defect-assisted non-radiative recombination and its effect on the green gap}

\author{Aurelien David}
\email{aurelien.david@polytechnique.org}
\affiliation{Soraa Inc., 6500 Kaiser Dr. Fremont CA 94555}
\author{Nathan G. Young}
\affiliation{Soraa Inc., 6500 Kaiser Dr. Fremont CA 94555}
\author{Christophe A. Hurni}
\affiliation{Soraa Inc., 6500 Kaiser Dr. Fremont CA 94555}
\author{Michael D. Craven}
\affiliation{Soraa Inc., 6500 Kaiser Dr. Fremont CA 94555}

\date{\today}

\begin{abstract}
Carrier lifetime measurements reveal that, contrary to common expectations, the high-current non-radiative recombination (droop) in III-Nitride light emitters is comprised of two contributions which scale with the cube of the carrier density: an intrinsic recombination --most likely standard Auger scattering-- and an extrinsic recombination which is proportional to the density of point defects. This second droop mechanism, which hasn't previously been observed, may be caused by an impurity-assisted Auger process. Further, it is shown that longer-wavelength emitters suffer from higher point defect recombinations, in turn causing an increase in the extrinsic droop process. It is proposed that this effect leads to the green gap, and that point defect reduction is a strategy to both vanquish the green gap and more generally improve quantum efficiency at high current.
\end{abstract}


\maketitle

Although modern III-Nitride light-emitting diodes (LEDs) can reach very high efficiency \cite{Hurni15}, they still suffer from two important limitations. First, their internal quantum efficiency (IQE) decreases at high current, an effect known as efficiency droop. It is by now generally believed that droop is, at least in part, caused by Auger scattering -- owing to the approximately-cubic dependence of the droop current on carrier density \cite{Shen07,David10a}, and the observation of Auger electrons \cite{Iveland13}. Second, IQE decreases at long wavelength (corresponding to a high In content in the active region), a phenomenon of controversial origin known as the green gap. Various explanations have been offered for the green gap. It has been proposed that high-In quantum wells (QWs) suffer from an increase in point defects \cite{Armstrong14,Hammersley15} -- however this should only impact low-current recombinations, whereas the green gap occurs at all currents. Other tentative explanations include an increase in electron-hole separation (leading to weaker radiative recombinations), due to increasing polarization fields and in-plane carrier localization \cite{Schulz15,Aufdermaur16,Nippert16b}. Crucially, these explanations don't clarify the role played by non-radiative recombinations in the green gap. The combined effects of droop and the green gap have made it challenging to achieve efficient long-wavelength LEDs at high current density, hindering important applications such as III-Nitride red-green-blue emitters.

In this Letter, we show that these two phenomena are connected and propose a mechanism for the green gap. We investigate droop dynamics in detail, and show that droop is in fact caused by a combination of two processes -- one intrinsic, the other extrinsic and scaling with point defect concentration. We then confirm that high In content leads to an increase in point defect recombinations, which in turn leads to an increase of the extrinsic droop component and the appearance of the green gap.

We study samples with 4 nm-thick single-QWs within p-i-n regions, grown on c-plane bulk GaN substrates by MOCVD; these samples include an InGaN underlayer (UL) beneath the p-i-n region, which improves material quality \cite{Akasaka04,Haller17,Haller18}. Importantly, the QW is placed at the center of the intrinsic region to avoid modulation-doping, which would alter the recombination dynamics \cite{Langer13}.

We first summarize important properties of the recombination dynamics in InGaN QWs. We measure carrier lifetimes ($\tau$) with an optical differential lifetime (ODL) technique under laser excitation, whose details are found in Ref.~\cite{David17a}. ODL uniquely enables the measurement of lifetimes down to low current density --in the Shockley-Read-Hall (SRH) regime-- and is devoid of electrical injection artifacts \cite{David16a}. We obtain $\tau$ as a function of the optical current density ($J$) in the QW. Concurrently, we measure the sample's absolute IQE as follows: the setup's geometry and the sample structure are designed to provide repeatable optical collection across samples, so that each sample's relative IQE is obtained directly from the luminescence intensity; absolute calibration is established by growing several pairs of samples, wherein one sample in the pair is used for ODL measurement and the other is processed as LEDs whose extraction efficiency is known \cite{David14a,Hurni15}, leading to an accurate IQE determination. These coupled measurements of lifetime and IQE yield the carrier density ($n$) in the QW and the radiative and non-radiative rates ($G_R$, $G_{NR}$), from which we derive three important quantities:

\begin{equation}
\label{Eq:abc}
a = G_{NR} / n,\ b = G_R / n^2,\ c = (G_{NR}-An)/n^3 .
\end{equation}

These quantities assume a simple interpretation in the framework of the well-known $ABC$ model: at low current, $a$ should be equal to the SRH coefficient $A$; while at all currents, $b$ and $c$ should be constant, and respectively equal to the radiative coefficient $B$ and the Auger coefficient $C$. For now, we place ourselves in this framework -- although the interpretation of $c$ will be refined hereafter. In practice $A$ is first extracted from the low-current plateau of $a$, and this value is then subtracted in Eq.~\ref{Eq:abc} to calculate $c$.

\begin{figure}[!!!tttth]
\includegraphics[width=8.5cm]{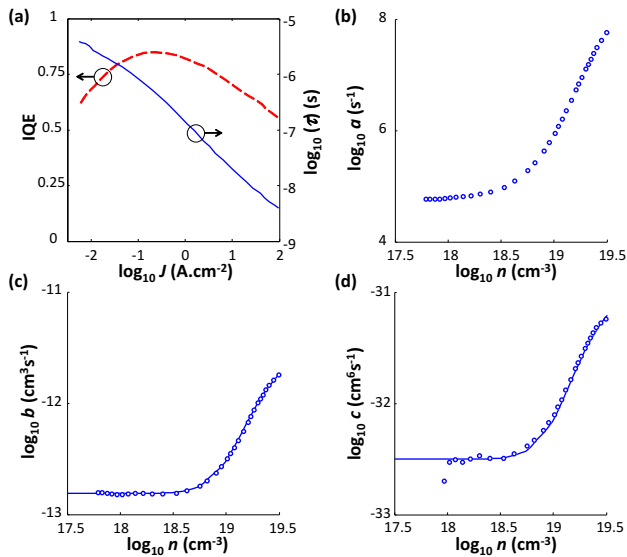}
\caption{Results from ODL measurements of a single-QW sample. (a) Lifetime (full line) and IQE (dashed line). (b) Non-radiative component $a$, whose low-current plateau gives the SRH coefficient $A$. (c) Radiative component $b$ showing the carrier-dependence of radiative recombinations. 
Symbols: experimental data; line: empirical fit, from which the screening function $S$ is derived (i.e. $b \sim  S^2$). (d) Non-radiative component $c$ showing the carrier-dependence of Auger scattering. Symbols: experimental data; line: fit by $c \sim S^p$.}
\label{Fig1}
\end{figure}

Experiments have shown that $b$ and $c$ are in fact current-dependent quantities \cite{David17a}. As shown in Fig.~\ref{Fig1} (for a sample with [In]=13\%), they increase with carrier density due to screening of the polarization field across the QW -- which induces an increase of the electron-hole overlap $I=\int\psi_e \psi_h$. It is well-known that $B \sim I^2$ \cite{Lasher64}. Regarding Auger scattering, experiments and calculations have shown that $C\sim I^p$, with $p \approx 2-3$ (depending on the mixture of $eeh$ and $ehh$ contributions) \cite{David10b,Jones17,David17a}.

$I$ can be decomposed according to $I(n)=I_0 S(n) $, where $I_0$ is the low-current value of $I$ and $S$ a screening function accounting for the effect of carriers on the potential, with $S$ normalized to unity at low current. Often $S$ is evaluated numerically from a simple one-dimensional Poisson-Schrodinger model, as was done in Ref.~\cite{David17a} -- however, such computations are of limited accuracy as they ignore in-plane carrier localization. Instead, in this Letter, we derive $S$ empirically from $S = \sqrt{b/B_0}$. The overlap dependence of $C$ can then be explicitly verified, as shown on Fig.~\ref{Fig1}(d): we find that $p \approx 2.4$, indicating a mixture of $eeh$ and $ehh$ Auger processes. Thus, we have $b = B_0 S^2$ and $c=C_0 S^p$, with the low-current values (subscript 0) obtained before the onset of screening. 

Further, we showed in Ref.~\cite{David17b} that the coefficient A, derived from Eq.~\ref{Eq:abc}, also scales with wavefunction overlap. Indeed, both electrons and holes need to be present at a point defect to complete a SRH recombination cycle. Experimentally, we find $A \sim I^q$ with $q \approx 1.6$ \footnote{This exponent is slightly different from the value $q=1.3$ we found in Ref.~\cite{David17b}. This is because the former exponent was derived from a 1D Poisson-Schrodinger model of $I$, whereas here we evaluate $I$ empirically from $I\sim \sqrt{B}$}. 

In summary, all three recombination mechanisms scale with similar exponents of the overlap $I$. Thus, for various active regions spanning a large range of overlaps, these dependencies cancel out, leading to nearly-constant peak IQE, as shown in Ref.~\cite{David17b}.

\begin{table}
\caption{\label{table} Sample details (UL refers to the relative thickness of the UL layer, values of $A$ and  $c_0$ are in log$_{10}$ scale).}
\begin{ruledtabular}
\begin{tabular}{c c c c c c}
\#		&	growth	&	UL &   $\Delta T$  &  $A$ $(s^{-1})$  &   $c_0$ $(cm^{6}s^{-1}) $   \\
\hline
1  &  1  &  100\%  &  0    &   4.78  &  -32.49  \\
2  &  2  &  100\%  &  0    &   5.44	 &  -32.10	\\
3  &  2  &  80\%   &  0    &   5.67	 &  -31.96	\\
4  &  2  &  50\%   &  0    &   6.38	 &  -31.33  \\
5  &  2  &  20\%   &  0    &   6.72	 &  -31.07  \\
6  &  2  &  0\%    &  0    &   7.21	 &  -30.79  \\
7  &  2  &  100\%  & +150C &   6.41	 &  -31.33	\\
\end{tabular}
\end{ruledtabular}
\label{Table1}
\end{table}

\begin{figure}[!!!thhhhhhhhhhhhb]
\includegraphics[width=8.5cm]{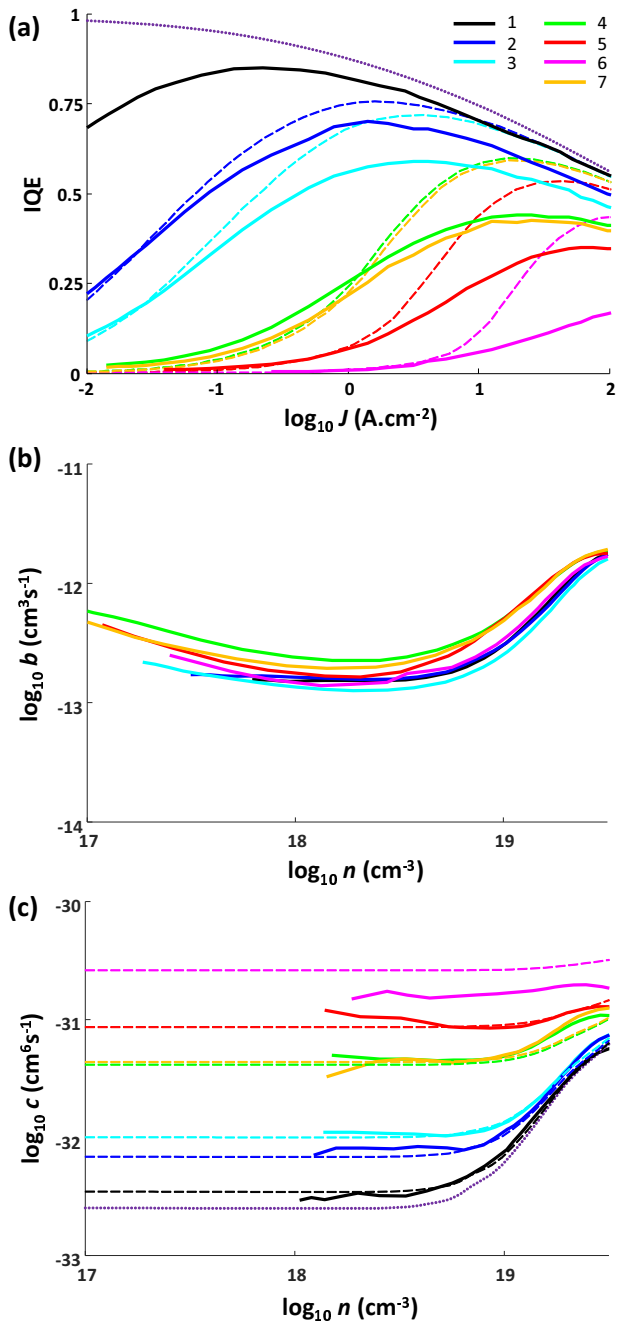}
\caption{ODL-derived results of Table~\ref{Table1} samples. (a) Full lines: experimental IQE. Dashed lines: modeled IQE for varying $A$ but constant values of $b$ and $c$; the model doesn't predict the decrease in IQE at high current. Dotted line: theoretical IQE limit in the intrinsic case ($A=0$, $D=0$). (b) The radiative term $b$ is nearly-identical for all samples. (c) The non-radiative term $c$ (full lines) varies across the series. Dashed lines: fits according to Eq.~\ref{Eq:intrinsic_extrinsic}. Dotted line: intrinsic Auger component, resulting in the dotted line in (a).}
\label{Fig2}
\end{figure}

Given this backdrop, we now study a series of samples where the SRH rate is intentionally varied by one of three methods. In a first case, we vary the epi growth between conditions labeled \textit{1} and \textit{2} (with \textit{2} being comparatively worse); in a second case, we vary the UL thickness (with a thinner UL leading to lower IQE); in a third case, we increase by $\Delta T$ the growth temperature of the intrinsic GaN layer underneath the QW \footnote{More precisely, the first half of this layer is grown hot; the second half is grown at normal temperature, to ensure good morphology before growing the QW} (leading to lower IQE). Importantly, in all samples, the active region is identical, with [In]$=13\%$. Table~\ref{Table1} shows the details of the samples. Fig.~\ref{Fig2} shows each sample's IQE and quantities ($b$, $c$) determined from ODL measurements. Quantity $a$ (not shown) displays a clear low-current plateau for each sample, from which $A$ is extracted. The fact that an increase in growth temperature in a layer well-below the active region leads to an increase in point defects (see sample 7), as previously noted in Ref.~\cite{Haller18}, suggests that the defect at play may be intrinsic rather than extrinsic.

IQE varies significantly across the series. The samples with higher $A$ suffer from low IQE at low current, as expected. However, their high-current IQE is also impacted. This is surprising because SRH recombinations should have little influence beyond $\sim 100 A.cm^{-2}$. To illustrate this, we model the expected impact of increasing $A$ while $b$ and $c$ remain unchanged for each sample (with values of $b$, $c$ taken from the brightest sample). Fig.~\ref{Fig2}(a) shows all modeled curves converging at high $J$, in contrast to the data. To explain this discrepancy, we turn our attention to the other recombination channels.

Fig.~\ref{Fig2}(b) shows that $b$ is near-identical for all samples, indicating the radiative rate is the same, as expected since they have the same active region. Incidentally, this confirms the accuracy of the ODL technique for measuring recombination rates, since all samples have markedly different lifetimes and IQEs, but the same $b$ is derived from each.

In contrast, in Fig.~\ref{Fig2}(c), $c$ shows a strong variation (more than one order of magnitude) across the series: the samples with higher $A$ also have higher $c$.  This is unexpected if we simply attribute $C$ to Auger scattering -- an intrinsic process, which should be constant for all samples just like radiative recombination. Instead, the interpretation of $c$ has to be revised; the data lends itself to the following decomposition:

\begin{equation}
c = C_0 S^p n^3 + D n^3 ,
\label{Eq:intrinsic_extrinsic}
\end{equation}

where $C_0$ is an intrinsic Auger coefficient (constant for all samples) and $D$ denotes a second, extrinsic droop process \textit{which scales with} $A$. This decomposition yields good fits for values of $C_0$ in the range 2-3E-33$~cm^{6}s^{-1}$. Hereafter, we use $C_0=$2.5E-33$~cm^{6}s^{-1}$, which leads to a close match to experimental data (shown in Fig.~\ref{Fig2}(c)), from which we derive the value of $D$ for each sample. It is worth noticing that the two processes of Eq.~\ref{Eq:intrinsic_extrinsic} differ in their wavefunction-overlap dependence (the intrinsic process increases at high density due to field screening, whereas the extrinsic process doesn't), which shows that the extrinsic process is not simply caused by an increase in magnitude of intrinsic Auger scattering.

\begin{figure}[!!!tttth]
\includegraphics[width=8.5cm]{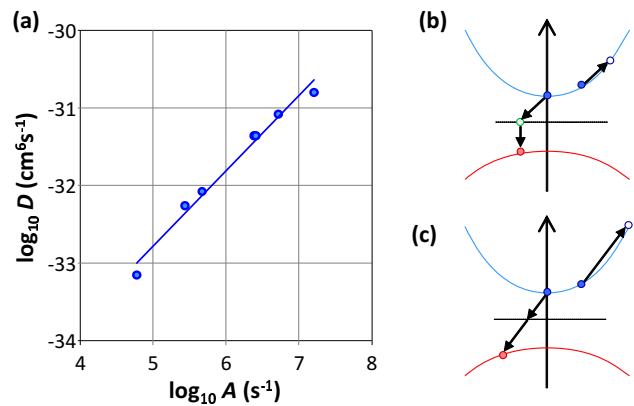}
\caption{(a) Correlation between the SRH coefficient $A$ and the extrinsic droop coefficient $D$ (in log-log scale), with a slope of $1 \pm 0.2$. (b, c) Band diagrams for possible high-order processes involving a trap: (b) electron Auger scattering into a trap followed by SRH hole recombination with the trap; (c) \textit{eeh} Auger process where the trap level acts as a virtual state.}
\label{Fig3}
\end{figure}

Fig.~\ref{Fig3} shows the resulting correlation between $A$ and $D$. The two coefficients are correlated with a slope in log-log scale of about unity (the precise value depends on the assumed value of $C_0$, but remains in the range 0.8-1.2). We conclude that to first order $D=k.A$ with $k=2$E-38~$cm^{6}$. This linear relationship suggests a causal relationship between defects and extrinsic droop. Namely, for a given active region design, $A$ is proportional to the concentration of a point defect causing SRH recombinations. $D$ scales \textit{linearly} with $A$, suggesting that this same point defect also induces an Auger-like non-radiative process, which adds to the intrinsic Auger scattering present in all samples. Only for sample 1, which has the lowest defect density, is the droop near the intrinsic limit indicated in Fig.~\ref{Fig2}. Crucially, this shows that defect reduction is essential for high-current performance. 

\begin{figure*}[!!!!!!!!!!!!!!!!!!!!!!!!!!!!!!thhhhhhhhhhhhb]
\includegraphics[width=\textwidth]{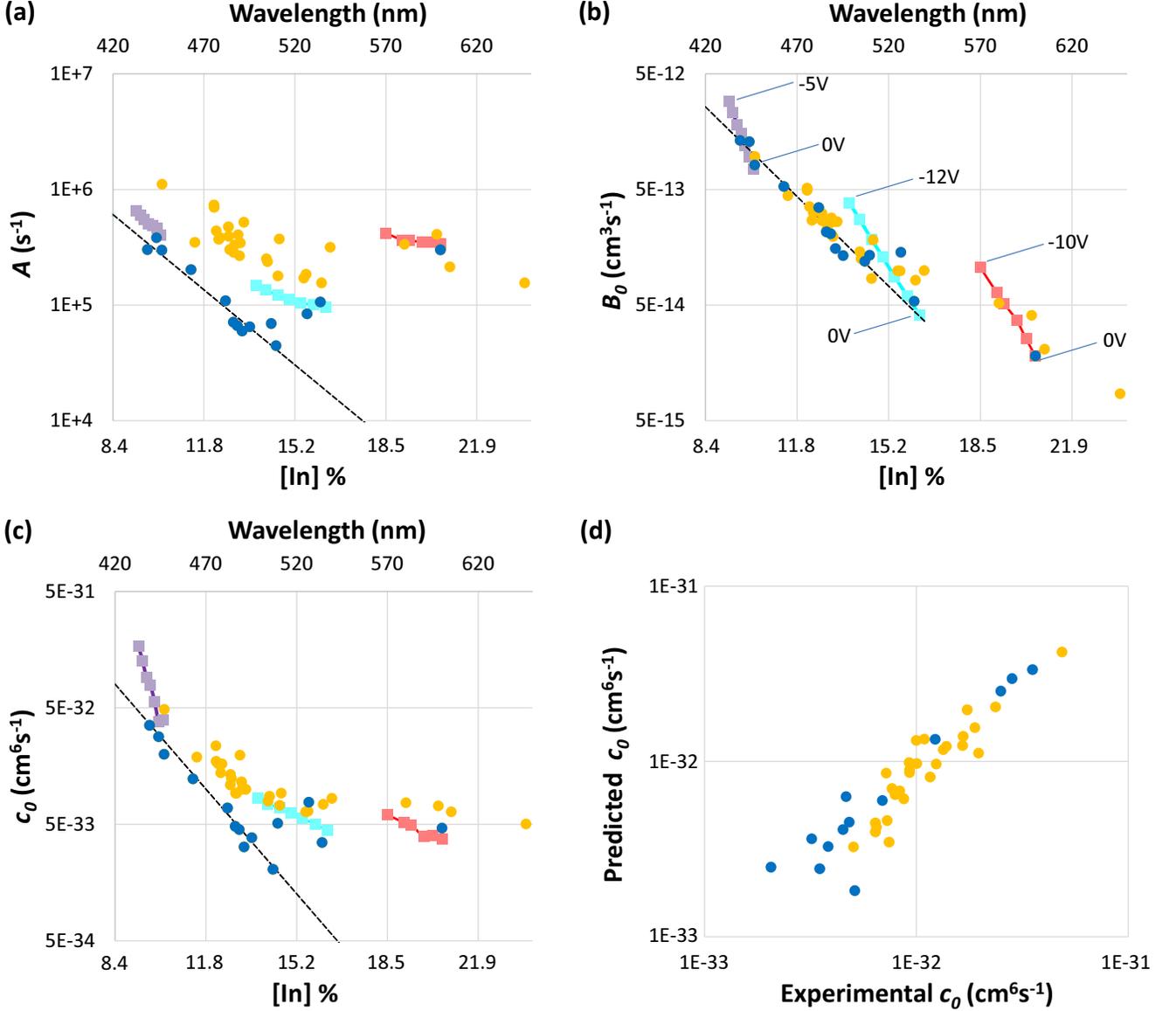}
\caption{(a-c) Recombination coefficients $A$, $B_0$ and $c_0$, plotted as a function of emission wavelength and corresponding [In]. Symbols represent the following. Dots: Samples of varying composition (blue dots: epi condition \textit{1}; yellow dots: epi condition \textit{2}). Dashed lines: empirical wavefunction dependences $A\sim I^q$ ($q=1.6$), $B\sim I^2$ and $C\sim I^p$ ($p=2.3$). Full lines with squares: samples under reverse bias (purple/cyan/red: [In]=10\%/16\%/22\% respectively). The bias endpoints are indicated on (b). (d) Correlation between measured value of $c_0$ and predicted value according to Eq.~\ref{Eq:intrinsic_extrinsic}. }
\label{Fig4}
\end{figure*}

Theoretical work has shown that the magnitude of the Auger coefficient in III-Nitrides can be explained by considering phonon- and disorder-assisted processes \cite{Kioupakis11,Kioupakis15}. Such microscopic processes are plausible explanations for the intrinsic droop current reported here. On the other hand, these processes don't involve defects, and cannot account for the extrinsic droop component. Rather, they should be constant across samples since the structure of the active region is unchanged.

Various high-order non-radiative mechanisms involving traps can be considered to account for the extrinsic process. However, its well-defined cubic dependence on the carrier density (manifested as clear plateaus across more than a decade in carrier density in Fig.~\ref{Fig2}(c)) imposes restrictions. Fig.~\ref{Fig3}(b) shows a process often considered in other materials: a combination of an Auger scattering to a trap state \cite{Landsberg64,Fossum83} followed by an SRH recombination; it can be verified that such processes have a carrier dependence in-between $n$ and $n^2$. More generally, processes generating a population in the trap state will deviate from an $n^3$ scaling. On the other hand, the mechanism of Fig.~\ref{Fig3}(c) (Auger scattering involving the trap level as a virtual state, without population buildup) would possess the proper carrier dependence, and is a possible candidate -- the magnitude of such a process in GaN hasn't been investigated. Qualitatively, defect states close to mid-gap (which are the likeliest source of SRH recombination) also reduce the energy threshold for Auger scattering, making such defect-assisted transitions plausible. Another possibility is enhancement of conventional Auger recombination by scattering at charged-defects -- although this was investigated theoretically in Ref.~\cite{Kioupakis15} and predicted to be negligible (even for high charged-defect densities precluded in our samples). Non-Auger processes might even be considered, such as the density-activated SRH mechanism of Ref.~\cite{Hader10} (however, that mechanism doesn’t have a prescribed dependence on carrier density, making it a less natural explanation than an Auger process). Further theoretical exploration would be warranted to determine which specific process may account for our observations. 

To explore the broader implications of this extrinsic droop process, we now consider samples of varying QW composition in the range [In]$ = 8-25\%$ -- i.e. spanning the violet-to-green range \footnote{In this study, the high-In samples emit at long wavelength at low current (see Fig.~\ref{Fig4}) due to the strong QW polarization fields in our test structures. At higher current, the wavelength shifts to a more typical value due to field screening.}. We grow series of such samples using epi conditions \textit{1} and \textit{2} mentioned above (and a full UL thickness of $100\%$). The study comprises 44 samples, whose values of $A$, $B_0$ and $c_0=C_0+D$ are measured by ODL and shown in Fig.~\ref{Fig4}. Their peak IQE is roughly constant in the violet-blue range (for a given epi condition) but decreases at longer wavelength -- a manifestation of the green gap. The radiative rate is well-behaved, with Fig.~\ref{Fig4}(b) showing that $B_0$ decreases with higher [In], as anticipated due to the decreasing wavefunction overlap and regardless of epi growth conditions.

Next we consider $A$ (Fig.~\ref{Fig4}(a)) -- which we expect to scale both with defect density and wavefunction overlap. Indeed, for moderate QW compositions ([In]$=8-15\%$), $A$ decreases with [In] -- the trend is clearest in samples with epi condition \textit{1}, but is also observed with condition \textit{2} (albeit with higher values of $A$ and more variability). This corresponds to the aforementioned regime, where the point defect incorporation remains constant and variations in $A$ are driven by wavefunction overlap (i.e. $A \sim I^q$). However, for higher [In], this trend breaks down and $A$ remains roughly constant as the QW composition increases. This marks a departure from the previous regime, and indicates that either \textit{(i)} the point defect density increases with [In] and/or \textit{(ii)} the overlap dependence of $A$ is altered.

Remarkably, the trend for the droop coefficient $c_0=C_0+D$ closely follows that of $A$. For a given growth condition, $c_0$ first decreases with overlap before reaching an approximate plateau. We propose that the departure of $c_0$ from the intrinsic regime $I^p$ is driven by variations in point defects, and hence $D$. To verify this, we show in Fig.~\ref{Fig4}(d) the values of $c_0$ predicted by Eq.~\ref{Eq:intrinsic_extrinsic} -- these closely match the experiment, confirming a systematic coupling between SRH defects and extrinsic droop.

To further understand the behavior of $A$ and $c_0$, we perform additional ODL measurements with a reverse bias applied to the samples. The bias reduces the electric field across the QW \cite{Schwarz07} and the resulting quantum-confined Stark effect, enabling a direct measurement of the overlap-dependence of recombinations in a given sample. We measure three samples with [In]=10\%, 16\%, 22\% (note that a reverse-bias photocarrier leakage current is observed for the 10\% sample \cite{David10c}; this effect is corrected-for in the data analysis). The results are included in Fig.~\ref{Fig4}, using the samples' wavelength to superimpose the data with the previous experiment. The bias-induced increase in overlap leads to a well-behaved increase in $B_0$ for all samples (the wavelength-dependence of these curves is slightly different from that of the previous experiment: this is because it is induced by a change in field, rather than composition). In contrast, the overlap-dependence of $A$ is most pronounced for the low-content sample and nearly nonexistent for the high-content sample. Accordingly, $c_0$ shows a similar behavior. 

Thus, as the QW's indium content increases, there is a transition from a ``well-behaved'' regime where all recombinations have pronounced overlap dependence (following the dashed lines in Fig.~\ref{Fig4}), to the green gap regime where $A$ and $c_0$ become nearly constant (overlap-independent). This transition might be due to the formation of a different SRH-causing defect, or to a change in the physics governing SRH recombinations in high-composition QWs. We hypothesize that increased carrier localization in high-content samples may contribute to these observations.

In summary, we have shown two crucial results for understanding the efficiency of III-Nitride LEDs. First, droop has two contributions. In addition to the known intrinsic Auger scattering in low-defect materials, SRH-causing defects also induce an additional, extrinsic Auger-like non-radiative process which dominates droop in high-defect materials -- in stark contrast with the common expectation that droop and defects are independent. Second, in long-wavelength samples with high In content, SRH recombinations become weakly dependent on wavefunction overlap and their value remains constant. Correspondingly, the extrinsic droop process remains constant while the radiative rate is reduced -- a trend which has been overlooked in previous studies focusing solely on radiative dynamics. We propose that this unfavorable balance of recombinations causes the green gap, and that improvements in long-wavelength III-Nitride emitters \textit{at all currents} may be enabled by a reduction in SRH defects and the associated droop process.

\bibliography{Biblio_These}

\end{document}